\documentclass[journal=jacsat,manuscript=article]{achemso}

\usepackage[version=3]{mhchem} 
\usepackage{mathrsfs,amsmath,amssymb,amsthm}
\usepackage{textcomp}

\author{Francisco de los Santos}
\email{fdlsant@ugr.es}
\affiliation{Departamento de Electromagnetismo y F{\'\i}sica de la
Materia, Universidad de Granada, Fuentenueva s/n, 18071 Granada,
Spain}
\author{Giancarlo Franzese}
\email{gfranzese@ub.edu}
\affiliation{Departamento de F\'isica Fundamental,
Universidad de Barcelona, Diagonal 645, 08028 Barcelona, Spain}

\title[Understanding diffusion and density anomaly in a coarse-grained
model for water confined between   hydrophobic walls]
{Understanding diffusion and density anomaly in a coarse-grained
model for water confined between   hydrophobic walls}

\begin{document}
\begin{abstract}
We study, by Monte Carlo simulations, a coarse-grained model of
a water monolayer between hydrophobic walls at  partial hydration,
with a wall-to-wall distance of about 0.5~nm.  We analyze how the
diffusion constant parallel to the walls, $D_{\parallel}$, changes and
correlates to the phase diagram of the system.
We find a locus of $D_{\parallel}$ maxima
 and a locus of $D_{\parallel}$ minima along isotherms, with lines of constant $D_{\parallel}$
 resembling the melting line of bulk water.  The two loci of $D_{\parallel}$ extrema
 envelope the line of temperatures of density maxima at constant $P$. We
 show how these loci are related to the anomalous volume behavior
 due to the hydrogen bonds. At much lower $T$, confined water
becomes subdiffusive, and we discuss how this behavior is a
consequence of the increased correlations among water molecules when the
hydrogen bond network develops.
 Within the subdiffusive region, although translations are largely
 hampered, we observe that the hydrogen  bond network can equilibrate  and its
 rearrangement is responsible for the appearance of density
 minima along isobars. We clarify that the minima are not necessarily
 related to the saturation of the hydrogen bond network.
\end{abstract}

\section{Introduction}

Water displays many thermodynamic and dynamic anomalies
\cite{debenedetti_stanley,life,footnote}.  Although the origin of
these anomalies can usually be traced back to the properties of
 hydrogen bonding between water molecules, their quantitative
understanding is a goal to which a great deal of effort is being
devoted. Experiments show that these anomalies are evident where
liquid water is stable, but are
stronger below the melting line where water is supercooled,
i.e. metastable with respect to crystal ice.
Bulk water can be kept in this supercooled state down to about 235 K
at 1 atm \cite{debenedetti-book}.  The experimental limit of stability
of supercooled water with respect to crystal ice defines the
homogeneous crystallization temperature $T_h(P)$, which reaches its
lowest value of 181 K at about 2000 atm.  In an attempt to
rationalize these anomalies, several ideas have been proposed,
including the stability limit (SL) conjecture
\cite{Speedy82}, the
liquid-liquid critical point  (LLCP) scenario \cite{llcp}, the
singularity-free (SF)
hypotheses
\cite{Sastry1996}, and the critical
point free (CPF) scenario
\cite{Angell2008}.  All these stimulating ideas are
consistent with the experimental properties of water, but hypothesize
different behaviors below $T_h(P)$.  Although these differences cannot
be directly tested in experiments, their implications in the
interpretation of water properties could be different, or could be
relevant for other anomalous liquids
\cite{Franzese:2001ea,vilaseca:084507,Vilaseca2011419}. It is
therefore worthwhile to test these hypothesis in theoretical models.
Many authors resort to computer simulations of detailed models of
water (see for example \cite{abascal2011} for a short updated
list). However, this approach, both for molecular dynamics (MD)
\cite{abascal2010} or Monte Carlo (MC) simulations \cite{liu:104508}, faces
the problem of large computational times near $T_h(P)$, because water
equilibration time increases in an exponential way for decreasing $T$.

An alternative approach, which we follow here, is to consider a
coarse-grained model of water that allows to perform, on the one hand, 
efficient MC simulations and, on the other hand, 
theoretical calculations. In particular, we focus on a model for a
water monolayer confined between hydrophobic walls
\cite{FS2002,FDLS2009}. The interest for this case comes from the fact
that, under appropriate conditions, the formation of ice can be avoided
in experiments with water under confinement
\cite{Bellissent95,Zhang09,Mancinelli10}.  We consider here only the
case of confinement between infinite hydrophobic walls, while other
kind of confinements, mimicking porous hydrophobic materials, have been
considered in other works \cite{Strekalova10}.

The coarse-grained model considered here allows to gain an insight into
the properties of the scenarios that have been proposed for
supercooled water.  In particular, it is possible to show that within
the framework of this model all the scenarios proposed for supercooled
water differ only in the
relative strength of the directional (covalent) and the many-body
(cooperative) component of the hydrogen bond (HB)
\cite{Stokely2010}. When the many-body HB component is strong and the
directional HB component is weak, the model recovers the CPF scenario,
and shows that it coincides with the SL case.  When the many-body HB
component is zero, the model reproduces the SF scenario for any finite
directional HB component. Finally, for intermediate values of the two
HB components, the model recovers the LLCP scenario, with the
liquid-liquid phase transition having a negative slope in the
pressure-temperature ($P-T$) phase diagram. The LLCP occurs at
positive or negative pressure depending of the relative strength of
the many-body HB component with respect to the directional HB
component \cite{Stokely2010}.  Direct experimental evaluation of the
relative strengths of the two HB components is not straightforward,
but indirect evaluations are consistent with values that, for the model,
would predict the LLCP scenario \cite{Stokely2010}.

\subsection{Dynamic properties}

Among the many dynamics anomalies of water, we will focus here on the
behavior of translational (self)diffusion constant $D$. In normal
liquids $D$ decreases when $P$ increases at constant $T$, while water
displays a large increase of $D$ in a delimited region of $P$-$T$
\cite{Jonas:1976uq,Prielmeier:1987kx}, e. g., with an enhancement of
about 60\% at 243~K when $P$ increases from 0.1~MPa to 150~MPa
\cite{Prielmeier:1987kx}.  The increase is observed up to about
200~MPa \cite{Prielmeier88,Cooperativity}.

Computer simulations of bulk water with detailed models (see
Ref.s~\cite{Reddy:1987bh,Gallo:1996dq,Starr:1999qf,Starr:1999ve,Netz:2001tg}
and references therein)
and lattice models (e. g. \cite{Girardi2007,Szortyka:2007ql,szortyka:104904})
can reproduce, at least qualitatively, the anomalous behavior of $D$.
By analyzing the microscopic structure of water molecules in the
region of the anomalous increase of $D$, several authors have proposed
a relation between the behavior of $D$ and the structure of water.
For example, Ref.~\cite{Scala:2000kl} relates $D$ to the configurational
entropy, Ref.~\cite{Errington01} associates the minima in $D$ with
a maximum in orientational order, and Ref.~\cite{Errington06} 
shows, for a water-like isotropic potential, the connection of
the anomaly in $D$, and in other quantities, with the density
dependence of the entropy in excess over the entropy of the ideal gas.
In particular, by classical molecular dynamic simulations
it has been observed that the increase of $P$ weakens the hydrogen
bonds, and thus increases $D$ \cite{Starr:1999qf,Starr:1999ve}.
This interpretation in terms of defects in the HB
network can be extended to negative $P$ \cite{Netz:2001tg}.  A similar
qualitative conclusion has been reached also by ab initio molecular
dynamics showing that $D$ is directly linked to network imperfections
\cite{Fernandez-Serra:2004eu}. Nevertheless, a quantitative relation
between the anomalous behavior of $D$ and the microscopic structure of
water is still missing.

In confinement, experiments show controversial results.  Without the
aim of reviewing the relevant literature, we recall here, as
examples, that the reduction of water $D$ can be of up to two orders
of magnitude between 260 K and 310 K, in hydrophilic NaX and NaA
zeolites \cite{Kamitakahara:2008eu}, or that the viscosity of water
between two hydrophilic surfaces, with $\leq 1$~nm interfacial
separation, is seven orders of magnitude greater than that of bulk water
at room temperature \cite{Major:2006ly}. Other experiments display
that $D$ decreases if the confinement increases, e. g. in MCM-41 with
pore radius between 1nm and 2nm
\cite{Takahara99,faraone:3963,Mallamace06,Sow-HsinChen08292006} or
MCM-48 with pore radius of about 1nm \cite{faraone:3963}, both slightly hydrophilic
due to the oxygen atoms in the silica structure, with the first
considered more hydrophobic than the second.  Similar results were
observed for channels of closed multiwalled hydrophobic carbon
nanotubes with diameter between 2nm and 5nm
\cite{Naguib:2004zr}. Nevertheless, other experiments
reveal an exceptionally fast mass transport for
 water
confined in carbon nanotubes of about 2nm \cite{Holt:2006vn} and 7nm
radius \cite{Majumder05}.

Computer simulations have been performed to rationalize the different
experiments. By MD  it has been found  a decrease of $D$ with
decrease of separation  between two hydrophilic surfaces at nanoscopic
distance  for SPC/E water \cite{Romero-Vargas-Castrillon:2009fk}.
Moreover, it has been shown for the same model that the decrease of
hydration largely decreases the mobility of
water molecules near the surface of MCM-41 or Vycor, consistent with the
interpretation that at low hydration the majority of them are bonded
to the surface  \cite{Gallo2010}.

In the case of hydrophobic confinement, the results are more controversial.
MD of TIP5P water nanoconfined between hydrophobic
smooth walls displays anomalous $D$ \cite{Kumar2005}, but only in the
direction parallel to the walls \cite{Han08}. $D$ results to be two orders of magnitude
smaller than in bulk and with the anomaly occurring in confinement at
lower $T$  than in bulk \cite{Kumar2005}. A similar large decrease of
mobility has been reported at ambient conditions for SPC/E water
between two large hydrophobic graphite-like plates
for separations below 1.3 nm \cite{Choudhury:2005ij}. Nevertheless,
first--principle MD simulations of the same model in similar conditions
show that the diffusion of water molecules become faster under
confinement, possibly due to weaker HBs at the interface
\cite{cicero2008}. A similar controversy is reported for simulations
of water in carbon nanotubes, with radii below 1~nm
\cite{cicero2008,Marti:2001kl,Mashl:2003tg,Liu:2005hc}.

\subsection{Our approach}

Our approach to the problem is to study by MC the local dynamics of a coarse-grained model
of confined water that will be defined in detail in the following.
The results described in the previous section for this model have been derived
by free-energy calculations within
mean--field approach and efficient MC simulations.  In particular, a
cluster MC dynamics allows to easily equilibrate the model at any $T$
and $P$ \cite{MSSSF2009}.  This specific MC approach is based on a
mapping of the thermodynamic model into a geometrical problem, using
an appropriate extension of the correlated-percolation approach
\cite{Cataudella1996,Franzese1996,FC1999,fbi2011}.  Nevertheless, as
an alternative it is possible to adopt a local MC algorithm
with the aim of studying the dynamic behavior of the model and compare
it with the experimental behavior of water \cite{KFS2008,KumarFS2008}.

This kind of study is useful for systems at equilibrium, although
approaching a glassy dynamics \cite{Franzese1998,Fierro1999,FC1999},
and makes the plausible assumption that, at a given $T$ and $P$, the
MC time step can be converted into real time units by a factor that does
not depend on time. This assumption is consistent with the
mode-coupling theory (MCT), according to which the long-time
relaxation ($\alpha$-relaxation) dynamics should be independent of the
microscopic dynamics \cite{Gotze:1992fk}, as tested in Lennard-Jones
mixtures \cite{Gleim}.  However, the comparison must be performed with
caution, because the time conversion factor for a given observable
could depend on both $T$ and $P$, when the interactions are
non-isotropic \cite{saw:164506}.  This dependence can be estimated by
comparing MC results with experiments, as done by Mazza et al. in
Ref. \cite{Mazza:2009} for the model under consideration here. The
comparison shows a linear relation between the logarithms of time in
real units and time in MC step at different $T$ and constant $P$
\cite{Mazza:2009}, corresponding to a power-law relation between the
two times. This relation can be
easily understood in the context of the present coarse-grained model, at
least at low $T$. At low enough $T$,
both  the experimental time and the MC time
follows a generalized Arrhenius relation \cite{Mazza:2009} where the characteristic
$T$-dependent activation energy corresponds to the energy needed to
break a HB.  The power-law relation between the two times
turns out to be a consequence of the choice of the model parameters
that implies a HB energy lower than in experiments. Therefore, by choosing appropriate model parameters the two
times would be directly proportional, with a proportionality factor
that would not depend on $T$ at constant $P$.

With this caveat in mind, we perform here an extensive MC
study of the thermodynamic and dynamic behavior of the coarse-grained
 model, presented in Ref. \cite{FS2002,FDLS2009}, for a
water monolayer confined between hydrophobic walls. In particular, we
consider the case in which water molecules can diffuse.
We identify the region where diffusion has an anomaly behavior,
finding maxima and minima of the diffusion coefficient at fixed $T$,
and we discuss how this anomaly is related to other anomalies of water.

In particular, thanks to the feature that the model can be tuned
from one scenario to another\cite{Stokely2010}, we can test if the diffusion
anomaly is related to some specific scenario for the
thermodynamics of the system at much lower $T$. Specifically, we
consider the LLCP and the SF scenarios. We do not observe any
relevant differences in the region where the system displays diffusion
anomaly.

We finally investigate the very low $T$ dynamics, observing a
subdiffusive regime and the approach to a glassy state at any
$P$. Under these conditions, we observe density minima, resembling
recent experiments \cite{Mallamace-density-min} and MD simulations
\cite{PhysRevLett.94.217802,Peter-H-Poole:2005ad}. In the present
model, the density minima appear in the vicinity of the glassy state
and as a consequence of the breaking of HBs to rearrange water
molecules for a better matching with the local order.
In particular, we observe that if the model does not take into account
the many-body interaction, the density minima are largely reduced.

\section{Coarse-grained model of a water monolayer}

We consider a water monolayer confined between two
hydrophobic smooth walls with  double periodic boundary conditions.
The walls are separated by a distance of about
$h=0.5$~nm. The system is considered at constant number $N$ of water
molecules, constant $T$ and constant $P$, leaving the volume $V$ free
to change. For a TIP5P-water monolayer confined between hydrophobic walls
it has been observed that, depending on the separation $h$,  water is
liquid or forms a quasi-two-dimensional square ice for
temperatures ranging between 260 K and 300 K, and negative  lateral
pressure \cite{PhysRevLett.91.025502,Kumar2005}. The square symmetry
is a consequence of the distortion
imposed by the walls to the tetrahedral HB network that would
otherwise form in bulk water.
Consistently with these findings, we divide the available volume $V$
into $\mathscr{N}$ square cells, each with volume $v=V/\mathscr{N}$,
square section of size $\sqrt{v/h}$ and height $h$, and
hydrate the system with $N\leq \mathscr{N}$ water molecules. 

In our coarse-graining we assume that each cell can at most host one
water molecule. Therefore, if $N= \mathscr{N}$, then each cell has one
molecule and the system is homogeneous in density. If $N<\mathscr{N}$,
some cells are empty. To each cell we associate an occupation variable
$n_i=0,1$ ($i=1,2,\ldots \mathscr{N}$) if it is vacant or occupied, respectively.
Having only one water molecule per cell between the
walls, for the sake of simplicity, we reduce the description of the
monolayer to a two-dimensional system.

The water-water interaction is decomposed into three terms. The first
accounts for all the isotropic contributions, including short-range
electronic orbitals repulsion and van der Waals attraction,
and is represented by a
Lennard-Jones potential truncated at a hard-core diameter $r_0$
\begin{equation}
U(r)\equiv
\begin{cases}
\infty & {\rm for} \ r\leq r_0 \cr
\displaystyle \epsilon \left[ \left( \frac{r_0}{r}\right)^{12}-\left( \frac{r_0}{r}\right)^6\right]
& {\rm for} \ r>r_0,\cr
\end{cases}
\label{LJ}
\end{equation}
where $\epsilon$ is the interaction energy and $r$ is the distance between two molecules.
In the coarse-grained representation, $r$ is the distance between the
center of occupied cells. The hard-core diameter $r_0$ is introduced
to simplify the implementation of the model and our tests show that
the results do not depend on its existence. We set $r_0\equiv 2.9$~\AA,
the water van der Waals diameter \cite{Narten67,Soper-Ricci-2000}, and
$\epsilon\equiv 5.8$~kJ/mol,
consistent with the value 5.5 kJ/mol of the estimate of the van der
Waals attraction based on isoelectronic molecules at optimal
separation  \cite{Henry2002}.

The second term of water-water interaction accounts for the
directional (covalent) component of the HB formation
\cite{Isaacs2000403,Pendas:2006bh}, with a characteristic energy that
we set $J\equiv 2.9$~kJ/mol. To account for the directionality
we adopt a geometrical definition in which the
HB breaks if
${\widehat{\rm OOH}}> 30^\circ$. Therefore, only 1/6 of the orientation
range $[0,360^\circ]$ in the OH--O plane is associated with a bonded
state.  We therefore associate to each molecule $i$ a bonding index
$\sigma_{ij}\in [1,2, \ldots q]$
with a discrete number of states $q$
describing the bonding state with a neighbor molecule $j$, and choose
$q\equiv 6$  to account for the entropy loss associated with the
formation of a HB. Due to the square symmetry, each
molecule has four neighbors and four bonding indices
$\sigma_{ij}$. Therefore, each molecule has $q^4=6^4=1296$ possible
bonding states. To form a HB between two molecules in two
occupied neighboring cells $i$ and $j$ (hence $n_i n_j=1$) we assume that the two
facing bonding indices $\sigma_{ij}$ and $\sigma_{ji}$ are in the same
state, i.e. $\delta_{\sigma_{ij},\sigma_{ji}}=1$, with
$\delta_{a,b}\equiv 1$ if $a=b$, and
$\delta_{a,b}\equiv 0$ otherwise.
Therefore, the directional term of the HB can be expressed
as
\begin{equation}
 H_J\equiv -J N_{HB},
\end{equation}
where
\begin{equation}
N_{HB}\equiv \sum_{\langle i, j  \rangle} n_i n_j
\delta_{\sigma_{ij},\sigma_{ji}}
\label{N_HB}
\end{equation}
is the number of HBs and the sum runs over all nearest-neighbor cells.

The directional interaction of HBs leads to local reduction
of density as a consequence of the reduced nearest neighbors with respect to close
molecular packing. This property is at the origin of the macroscopic density
maximum that occurs at $4^\circ$C at ambient pressure, a temperature
below which the number of HBs per molecule is about 3
\cite{Suresh00}. The effect can be observed in experiments mainly as a
change in the structure of the second shell of water molecules
\cite{Soper-Ricci-2000,Soper:2000jo,Soper08,bernabei:021505}.
Nevertheless,
to include the effect in a tractable way in the coarse-grained model,
we follow Ref.\cite{Sastry1996} and
consider that each formed HB leads to a small increase
of volume $v_{HB}$ per molecule, where $v_{HB}/v_0 \equiv 0.5$ is the
average density increase from low density ice Ih to high density ices
VI and VIII and $v_0\equiv hr_0^2$ is our approximation to the van der
Waals volume of a molecule.  Therefore, the total volume occupied by
water is
\begin{equation}
V_w\equiv Nv+N_{HB}v_{HB}.
\label{Vw}
\end{equation}
Note that the increase $v_{HB}$
corresponds to a larger volume per molecule, but not to a larger
separation $r$ between molecules, hence it does not affect the radial
term in \ref{LJ}.

The last term we include in the water interaction energy is a
many-body (cooperative) interaction among HBs, that favors
specific values of the probability distribution of O--O--O angles
\cite{Soper08,bernabei:021505} (see also Ref.~\cite{Cooperativity}
for a brief description of the quantum origin of the cooperative
interaction). Furthermore, the probability distribution of O--O--O
changes when comparing bulk and confined water \cite{Ricci09},
showing the disappearing of the fifth interstitial neighbor in the
confined case, and a shift of the maximum of the distribution toward
$90^\circ$ at low $T$, consistent with the symmetry chosen for the
coarse-graining in our model.
To account for this cooperative interaction, we include in the model
the term
\begin{equation}
 H_{\rm{Coop}}=-J_{\sigma}\sum_i n_i \sum_{(k,l)_i}
 \delta_{\sigma_{ik},\sigma_{il}},
\label{coop}
\end{equation}
where $(k,l)_i$ stands for the six different pairs of four bonding indices
of a molecule $i$, and $J_{\sigma}$ is related to the energy gain when
the bonding indices order in the same state, with the maximum gain
$-6J_{\sigma}$ per molecule, corresponding to the fully ordered case.

As shown by Stokely et al. \cite{Stokely2010}, by setting the
parameters $\epsilon$, $J$ and $v_{HB}$ to finite values and tuning
the parameter $J_{\sigma}$, it is possible to reproduce all the
scenarios that have been proposed for supercooled water. Here we
consider two cases: (i) $J_{\sigma}=0.29$ kJ/mol, corresponding to the
LLCP scenario, and (ii) $J_{\sigma}=0$,  corresponding to the SF
scenario.

The choice in (i) is consistent with the
experimental measure of  HBs in ice Ih, approximately 3 kJ/mol
stronger than in liquid water \cite{newref}.
If we entirely attribute this increase  to the cooperative component of the HB
\cite{Heggie1996}, we find $J_{\sigma}\simeq 0.25$ kJ/mol for the  two molecules forming
the HB  (each with an energy gain in absolute value $6J_{\sigma}$), which is consistent with our
choice.

The water-wall interaction is represented by a hard-core
exclusion. Although the interaction of water with a hydrophobic, infinitely
large, object could have a small attractive van der Waals component
or a soft repulsion, for the sake of simplicity we assume here that the
main effect of the confining walls is to inhibit the formation of ice,
as observed by  Zangi and Mark for $h<0.51$ nm
\cite{PhysRevLett.91.025502}.

Therefore, the enthalpy of the system at pressure $P$ is
\begin{equation}
H\equiv U(r)+H_J+H_{\rm{Coop}}+PV_w.
\label{H}
\end{equation}
For a given  occupancy ratio $N/\mathscr{N}$,
the state of the system is fully specified by the average number density
$N/V$ and the set of $\sigma_{ij}$.

\section{The Monte Carlo method}

We perform MC simulations in the $NPT$ ensemble for a system
partitioned into $\mathscr{N}=2500$ square cells and an occupancy ratio
$N/\mathscr{N}=0.75\%$, corresponding to $N=1875$ water molecules.
Since we allow for changes of the volume in the direction parallel to the
walls, the control parameter $P$ represents the pressure parallel to
the walls.
To test for size effects, we consider also $\mathscr{N}=400$ and
$\mathscr{N}=1600$ at the same $75\%$ occupancy ratio, with
$N=300$ and $N=1200$ respectively.
Our results do not show any  appreciable size effect among
these three cases.
Likewise, we do not find any significant differences for occupancy ratios
between $75\%$ and $90\%$.

MC and mean-field results for this model at full
occupancy
\cite{FYS2000,FS2002,FS-PhysA2002,FMS2003,FS2005,FS2007,KumarFBS2008,FranzeseSCKMCS2008,KFS2008,Mazza:2008,KumarFS2008}
were reported in previous works
\cite{MSSSF2009,Mazza:2009,dlSF2009,FDLS2009,Franzese2010,Odessa2009,Stokely2010,fbi2011,Strekalova10}.
An analysis of the differences between partial and full occupancy will
be presented elsewhere.

To generate equilibrium configurations we pick at random
a cell $i\in \{1,\ldots ,N\}$ and an integer $j \in \{0,1,2,3,4\}$.
If $j=0$, we choose at random one of the four neighboring cells of the cell $i$
and, if empty, we displace into it the molecule $i$ with probability
given by the following \ref{metro} of a Metropolis algorithm.

If $j \geq 1$ we set the bonding index $\sigma_{ij}$ to
any of the $q$ possible states, independent of its original state, and
accept the change with probability in  \ref{metro}.
One MC step consists of $5N$ of these trials followed by a
volume-change attempt, in which we select a
new volume at random in the interval $[V-\delta V, V+\delta V]$ with
$\delta V=0.5v_0$, and accept the move with probability in
\ref{metro}.
Since we change the volume $V$ in a continuous way, the volume per cell
$v$ and the distance between the center of cells $r$ also change in a
continuous way, as in an off-lattice model, despite the fact that the 
model has a fixed maximum number of
nearest neighbors, equal to four, due to the square lattice symmetry
adopted in the coarse-graining.

From the new and old configurations we calculate
$\Delta H\equiv H^{\rm {new}}-H^{\rm {old}}$ from \ref{H}, and
$\Delta S\equiv -Nk_B\ln(V^{\rm {new}}/V^{\rm {old}})$, where $k_B$ is
the Boltzmann constant. We accept the
new configuration with probability
\begin{equation}
\rm{min}\{1,\exp[-\beta(\Delta H -T\Delta S)]\},
\label{metro}
\end{equation}
where $\beta=1/(k_BT)$.

Following Ref.~\cite{Mazza:2009} we adopt real units to represent our
results.
The transformations  to real units
are based on rescaling the MC $T$, $P$, and $t$ from experimental data
for a monolayer of water adsorbed on lysozyme powder
\cite{Mazza:2009} and adjusting the MC $\rho$ to experimental values
around the maximum density at ambient pressure in a self-consistent way.
These transformations are meant to give
the order of magnitude of the calculated quantities.

In the calculation of  the Lennard-Jones interaction energy in
\ref{LJ} we test that there is no appreciable difference if we
introduce a cut-off at the 9th neighbor for the maximum
interaction range. To allow for a better equilibration of the system,
we follow an annealing protocol along isobars, starting at high $T$,
with $\mathscr{N}$ cells randomly occupied by $N$
water molecules, each molecule having a random
configuration of bonding indices, and with the total volume
$V=\mathscr{N}v_0$.  We equilibrate each state point for 0.2~ms
and produce data for 15~ms.

We calculate the coefficient $D_{\parallel}$ of water diffusion
parallel to the plates from the Einstein relation in two dimensions
\begin{equation}
D_{\parallel}\equiv \lim_{t\to \infty} \frac{\langle |{\bf r}_i(t+t_0)-{\bf r}_i(t)|^2 \rangle}{4t}
\label{diffusion_coeff}
\end{equation}
where ${\bf r}_i(t)$ denotes the projection onto the plates of the
position of molecule $i$ at time $t$, and $\langle \cdot \rangle$
stands for the average over all molecules $i$  and over different
values of $t_0$.

To avoid correlations in the calculations of $D_{\parallel}$ and all
the other quantities, we perform averages over blocks of
$\tau_0=0.8~\mu$s, sampled every 80~ps. We check
that $\tau_0$ approximately corresponds to the time needed for a molecule
to reach its image points for most of the temperatures investigated.
However, for the lowest temperatures, we use up to $10 \tau_0$ as
block-length for averages.

\begin{figure}[ht!]
\includegraphics[width=12cm,angle=-90]{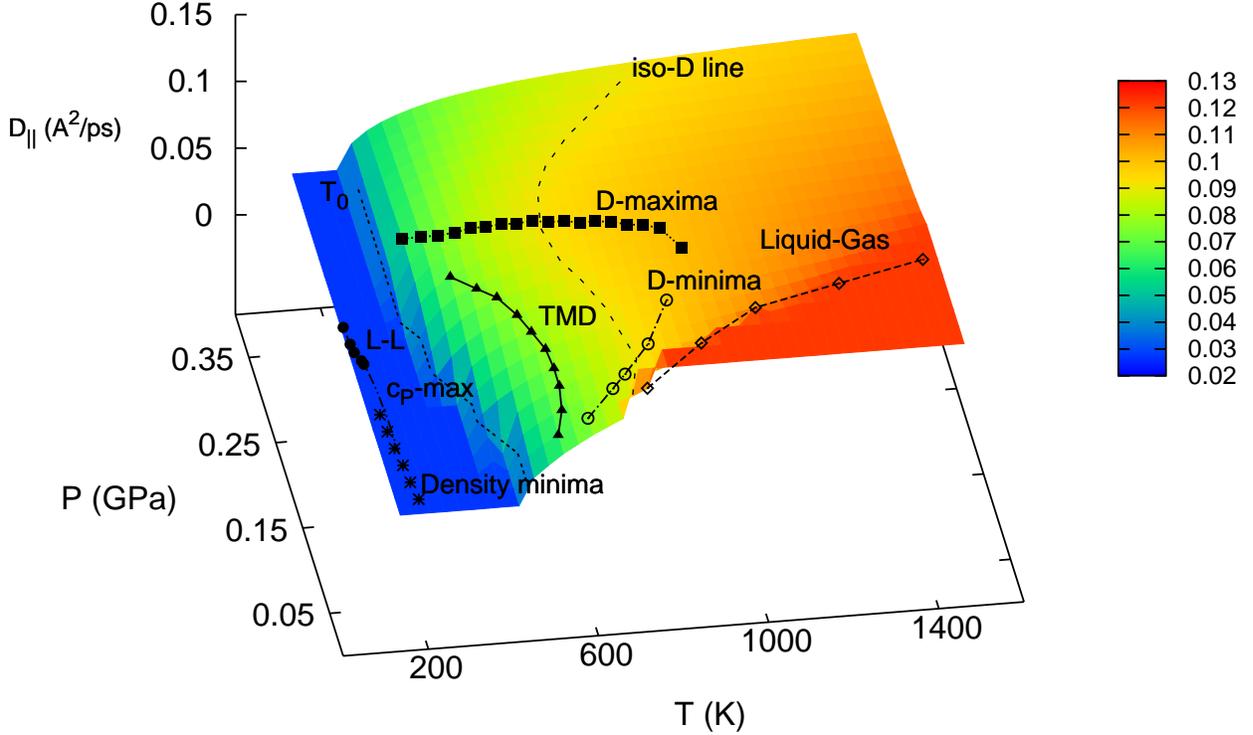}
\caption{$P-T$ phase diagram of a water monolayer nanoconfined between
  hydrophobic plates in which we emphasize how the diffusion constant $D_{\parallel}$
($z$-axis and color scale) changes in relation with the thermodynamic behavior.
The liquid-gas phase transition line (open diamonds) ends in the
liquid-gas critical point (diamond symbol at highest $T$).
The loci of isothermal maxima of $D_{\parallel}$, $D_{\parallel}^{\rm{MAX}}$
(solid squares), and minima, $D_{\parallel}^{\rm{min}}$ (open
circles), include the TMD line (solid triangles).
Loci at constant $D_{\parallel}$ (e.g. the dashed line
marked as `Iso-$D$') resemble in their reentrant
behavior  the water melting line.
$D_{\parallel}$ values above 0.12 \AA/ps$^2$ (gas phase) and below
0.03 \AA/ps$^2$ are not
color-graded. State points with $D_{\parallel}<$0.03 \AA/ps$^2$ are
below  the onset $T_o(P)$ (dotted line) of the
subdiffusive regime and their diffusion coefficient cannot be defined.
For $T<T_o(P)$, we
observe a locus of density minima (asterisks), a locus of
discontinuity in density (solid circles), a locus of maxima of $C_P$
(dot-dashed line) as in a liquid-liquid phase transition
ending in a critical point.
Here we use  $\epsilon=5.8$~kJ/mol, $J=2.9$~kJ/mol, $J_\sigma=0.29$~kJ/mol, $v_{HB}/
v_0=0.5$ as parameters for the coarse-grained water model.}
\label{phase-diagram}
\end{figure}

\section{Results and discussion}

\subsection{Phase diagram}

At high $T$ we find the gas phase, separated from the liquid phase by a
first-order phase transition ending in a critical point
(\ref{phase-diagram}). The liquid-gas critical point for the
hydrophobically confined monolayer
occurs at a pressure and temperature that are
higher than that of bulk water, qualitatively consistent with the
results of MD simulations for TIP4P water in
hydrophobic confinement \cite{Gallo07}.

\begin{figure}[]
\includegraphics[width=12cm,angle=-90]{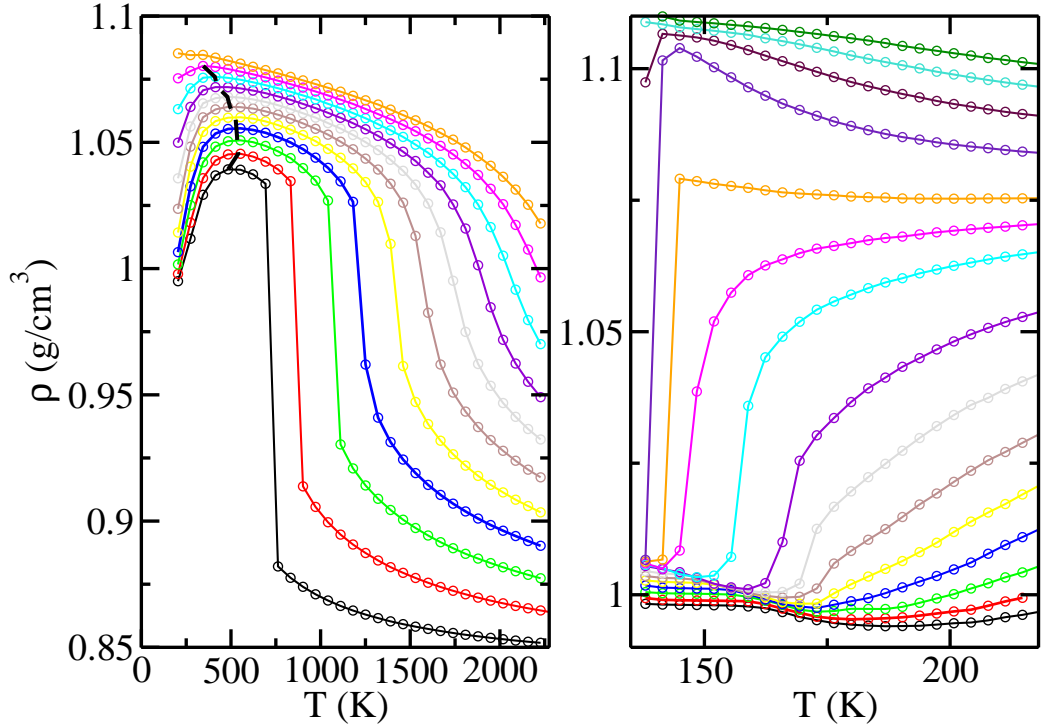}
\caption{Density $\rho$ as a function of temperature $T$ along isobars
 at $0.02$~GPa $\leq P \leq 0.22$~GPa separated by 0.02~GPa increments,
 as calculated by MC annealing.
(Left panel) At low $P$, the discontinuity in $\rho$ marks the
gas-to-liquid spinodal temperature $T_S^{\rm{LG}}(P)$, vanishing at
higher $P$ into the liquid-gas critical point. The dashed line marks
the temperature $T_{\rm MD}(P)$ of maximum density. For clarity we
show only a selection of the simulated state points.
(Right panel) Enlarged view at low $T$ along isobars at
$0.02$~GPa $\leq P \leq 0.30$~GPa separated by 0.02~GPa increments.
Model parameters as in \ref{phase-diagram}.}
\label{density}
\end{figure}

By annealing the system from high temperatures,
we find a discontinuous change in density along isobars
(\ref{density}). This change occurs at the spinodal temperature
$T_S^{\rm{LG}}(P)$, that marks the limit of stability of the gas phase with
respect to the liquid phase. The change is very large at low $P$ and
vanishes as the liquid-gas critical point is approached. At the
critical point, by definition, the isobar has infinite negative slope.
At $P$ higher than the critical pressure, the minimum slope of
$\rho(T)$ become finite and decreases in absolute value for increasing
$P$. The locus of these minima corresponds to the locus of maxima of the
isobaric thermal expansion coefficient  $\alpha_P^{\rm{max}}(P)\equiv
\rm{min_T}\{-\partial \ln
(\rho v_0)/\partial T|_P\}$, with the maxima  decreasing in value by increasing $P$.

By changing the simulation protocol across $T_S^{\rm{LG}}(P)$,
i.e. heating the system instead of annealing it, we find the typical
hysteresis associated to a first-order phase transition, with the
hysteresis vanishing as the liquid-gas critical point is approached.
All these results are consistent with previous findings for this
model~\cite{FYS2000,FS2002,FS-PhysA2002,FMS2003,FS2005,FS2007,KumarFBS2008,FranzeseSCKMCS2008,KFS2008,Mazza:2008,KumarFS2008,MSSSF2009,Mazza:2009,dlSF2009,FDLS2009,Franzese2010,Odessa2009,Stokely2010,fbi2011,Strekalova10}.
Here we add the information about the diffusion constant.

\subsection{Diffusion maxima and minima}

Upon crossing the liquid-gas phase transition, the diffusion constant
$D_{\parallel}$ displays a discontinuous change that vanishes as the
critical point is approached. Above the critical point, $D_{\parallel}$
changes in a continuous way, as expected in a one-phase region.

\begin{figure}[ht!]
\includegraphics[width=12cm,angle=-90]{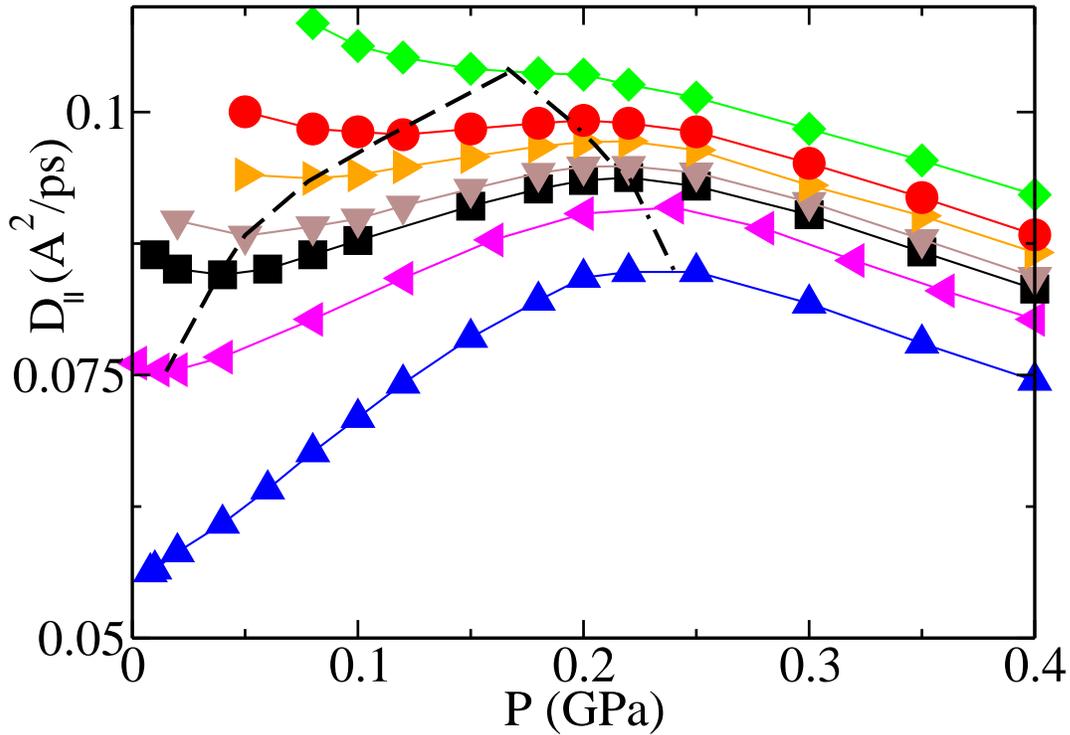}
\caption{Diffusion coefficient $D_{\parallel}$ as a function of
pressure along isotherms for (from bottom to top) $T=$483~K, 588~K,
658~K, 693~K, 762~K, 832~K and 1006~K. For $T>832$~K, $D_{\parallel}$
is monotonic with $P$, while it has maxima $D_{\parallel}^{\rm{MAX}}$
and minima $D_{\parallel}^{\rm{min}}$ for $588$~K$\leq T\leq 832$~K,
and a maximum for $T=$483~K. The pressure $P^{\rm{D-MAX}}$ for
$D_{\parallel}^{\rm{MAX}}$ (dot-dashed line) decreases for increasing
$T$ and converges to the pressure $P^{\rm{D-min}}$ for $D_{\parallel}^{\rm{min}}$
(dashed line). $P^{\rm{D-min}}$ decreases for decreasing
$T$ and, possibly, becomes negative.
Model parameters as in \ref{phase-diagram}.
}
\label{diffusivity}
\end{figure}

In the liquid phase for $T>832$~K, $D_{\parallel}$ decreases
monotonically for increasing $P$, but has an anomalous non-monotonic
behavior at lower $T$ (\ref{diffusivity}). This behavior resembles the
known diffusion anomaly of bulk water. In particular, we find that $D_{\parallel}$
decreases below a maximum $D_{\parallel}^{\rm{MAX}}$ for decreasing
pressure at $P\lesssim 0.2$~GPa, as observed in bulk water
\cite{Prielmeier88,Cooperativity}.  The pressure at which we find
$D_{\parallel}^{\rm{MAX}}$, $P^{\rm{D-MAX}}$,  increases with decreasing $T$. At lower
$P$, $D_{\parallel}$ reaches a minimum $D_{\parallel}^{\rm{min}}$ at
a pressure $P^{\rm{D-min}}$ that decreases with decreasing $T$ and eventually becomes
negative for $T<588$~K (\ref{phase-diagram}).

As a consequence of the occurrence of $D_{\parallel}^{\rm{MAX}}$ and
$D_{\parallel}^{\rm{min}}$, the state points with the same value of
$D_{\parallel}$ forms lines in the $P$-$T$ phase diagram that are not
monotonic as a function of $P$. Therefore, these lines at constant
$D_{\parallel}$, or iso-$D_{\parallel}$ lines (\ref{phase-diagram}),
have a positive slope in $P$-$T$ phase diagram for
$P>P^{\rm{D-MAX}}$ and for  $P<P^{\rm{D-min}}$, but a negative slope for
$P^{\rm{D-MAX}}>P>P^{\rm{D-min}}$.
It is interesting to observe that this change of
slopes in $P$-$T$ plane resembles the change of slope of the bulk
water melting line, at least for $P>P^{\rm{D-min}}$. Although the
present model does not include any representation
for the crystal, because we make the hypothesis that crystallization
is avoided under the conditions considered here, our finding of the change of slope
of the iso-$D_{\parallel}$ lines suggests that the shape of the melting line in
nanoconfinement could resemble that of bulk water. Moreover,
our observation hints that its shape would be mainly
determined by the slowing down of the dynamics.

Finally we observe that  the temperatures of $D_{\parallel}^{\rm MAX}$ and
$D_{\parallel}^{\rm min}$ are higher than those
found in confinement and bulk with MD simulations of TIP5P-water
\cite{Kumar2005}. This difference could be consistent with the higher
values observed here for the liquid-gas spinodal line. On the other
hand, the pressure at which we observe the onset of
$D_{\parallel}^{\rm MAX}$  is consistent with experimental results for
bulk water \cite{Prielmeier88,Cooperativity}.

\subsection{Density maxima and expansivity minima}

By decreasing $T$ along isobars, we find maxima in density at
a temperature $T_{\rm MD}(P)$ for $P\lesssim 0.2$~GPa, as observed in bulk water
\cite{Prielmeier88,Cooperativity} (\ref{density}). Above $P\simeq
0.2$~GPa the density increases regularly for decreasing $T$, while at
lower $P$ and moderate temperature $T<T_{\rm  MD}(P)$, the density of
liquid water decreases for decreasing $T$. As a result of this
anomalous density behavior the isobaric thermal expansion coefficient
(or expansivity)
$\alpha_P(T)\equiv -\partial \ln (\rho v_0)/\partial T|_P$
becomes negative for $T<T_{\rm  MD}(P)$, as
observed in bulk water \cite{debenedetti_stanley}.

The temperature $T_{\rm MD}(P)$ of maximum density  has, in the $P$-$T$ phase
diagram, a shape that resembles the one observed in bulk and in
nanoconfined water \cite{Kumar2005}, reaching a maximum temperature
at about 50~MPa, similar to what is found in the TIP5P water
model \cite{Kumar2005}.
As expected for water \cite{Errington01}, we find the $T_{\rm MD}$ locus
within the region where the diffusion anomaly occurs, delimited between
$P^{\rm{D-MAX}}(T)$ and $P^{\rm{D-min}}(T)$ (\ref{phase-diagram}).

For $T<T_{\rm MD}(P)$ we find that  the slope of
$\rho(T)$ along isobars increases by decreasing $T$ (\ref{density}),
implying a more pronounced negative
value for $\alpha_P$, consistent with experiments for bulk
\cite{Sorensen:1983lh}  and supercooled confined water
\cite{hare1986,Mallamace-density-min}. The slope decreases at lower
$T$, consistent with the occurrence of a minima in $\alpha_P$,
expected below 240~K from experiments in bulk water at ambient pressure
\cite{fuentevilla:195702}.

\begin{figure}
\includegraphics[width=12cm,angle=-90]{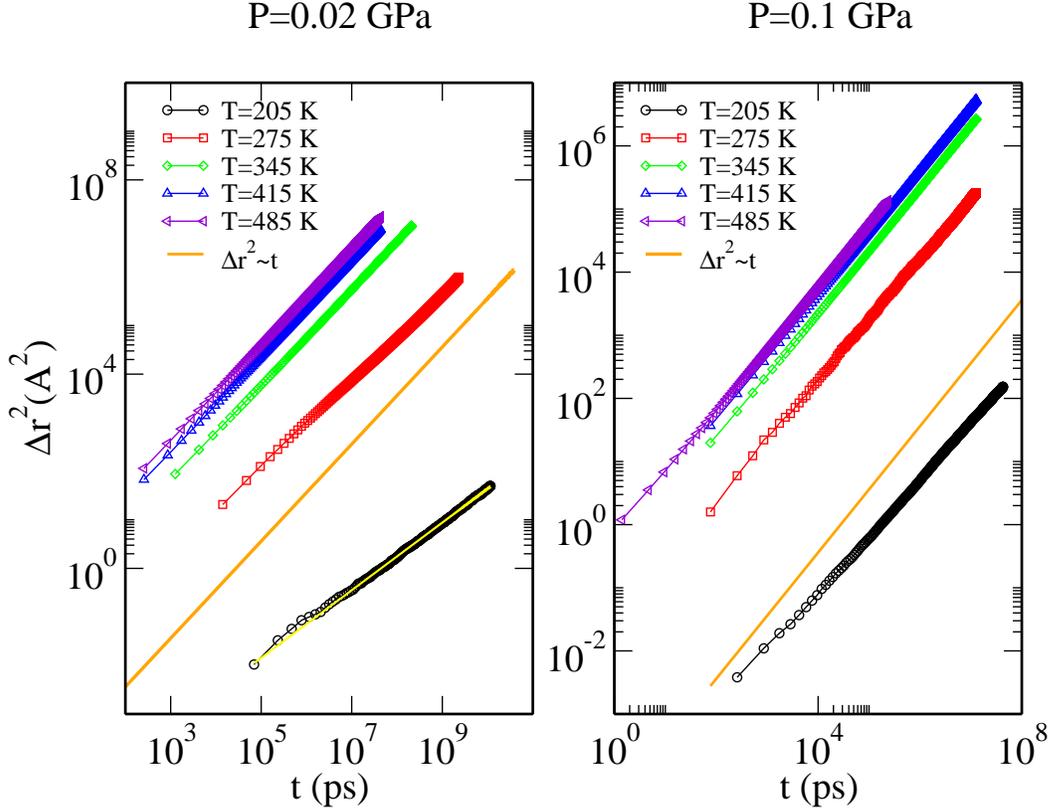}
\caption{Mean square displacement $\Delta r^2\equiv \langle |{\bf
  r}_i(t+t_0)-{\bf r}_i(t)|^2  \rangle$ as function of time $t$. In
log-log plot, the diffusive regime  $\Delta r^2\sim t$ has
a slope $1$ (parallel to the line without symbols). We calculate the
diffusion coefficient  $D_{\parallel}$ from  \ref{diffusion_coeff}
within the diffusive regime.
(Left panel) At $P=0.02$~GPa, for $T\leq 275$~K we find that the system does not
reach the diffusive regime and the long-time behavior ($t>10^6$~ps) is well
described by the subdiffusive relation $\Delta r^2\sim t^\alpha$, with
$\alpha=0.7$.
(Right Panel) At higher pressure $P=0.1$~GPa the onset of subdiffusive regime
occurs at lower $T$ with respect to the $P=0.02$~GPa case, for $T\leq
205$~K.
Model parameters as in \ref{phase-diagram}.
}
\label{subdiffusivity}
\end{figure}

\subsection{Subdiffusion}

Before describing in detail our findings about the density at lower
$T$, we observe that for $T\leq 275$~K at low $P=0.02$~GPa, within our
simulation time $\simeq 0.01$~s, the system does not reach the
diffusive regime, i.e. the mean square displacement $\langle |{\bf
r}_i(t+t_0)-{\bf r}_i(t)|^2 \rangle$ is no longer proportional to the time
$t$ (\ref{subdiffusivity}). We find that the long-time behavior
($t>10^6$~ps) is well described by the subdiffusive relation $\Delta
r^2\sim t^\alpha$, with $\alpha=0.7$.

Subdiffusive behavior is observed in experiments for water hydrating
mygloblin, at a hydration level corresponding to approximately one water
monolayer \cite{Settles-Doster1996}. The experiments show
water subdiffusion at 320 and 300~K,
with an exponent $\alpha=0.4$ \cite{Settles-Doster1996}.  This
subdiffusive behavior (also called ``anomalous diffusion'') has been
rationalized by several authors by means of simulations of
water, both in inorganic \cite{Gallo:2003vn,Gallo2010} and organic
confinement
\cite{Bizzarri:1996uq,Bizzarri:1996fk,Rocchi:1998kx,Oleinikova:2007bh},
with $\alpha$ exponents varying between 0.96 \cite{Bizzarri:1996fk}
and $0.45\pm 0.05$ \cite{Gallo:2003vn}. The proposed rationale is
that the subdiffusive behavior is due to the heterogeneity of the
surface and of the water-surface interaction. Nevertheless, this
interpretation does not apply to our case, where the surface is by
definition homogeneous and flat and the water-surface interaction is
only due to volume exclusion. In our case the subdiffusive dynamics is,
instead, originated by the increasing correlation among the water
molecules that will be discussed in the following subsections.

At higher pressure, $P=0.1$~GPa, we find that  the onset $T_o(P)$ of the
subdiffusive regime occurs at about $T\leq 205$~K, i.e. at lower $T$
with respect to $P=0.02$~GPa (\ref{subdiffusivity}). Therefore, within
this range of $P$, the temperature $T_o(P)$ is correlated
with $T_{\rm{iso-D}}(P)$ of iso-$D_{\parallel}$ lines,
being $T_{\rm{iso-D}}(P)-T_o(P)\simeq$ constant. This is no longer
true at $P>0.22$~GPa (\ref{phase-diagram}) and can be
understood in the framework of this model, because, as we will show in the
next subsections, for $P>0.22$~GPa the number of HBs vanishes at
low $T$
(\ref{nhb}).

\subsection{Density minima: relation with the cooperativity
  and the slow dynamics of the HB network}

\begin{figure}
\includegraphics[width=6cm,angle=-90]{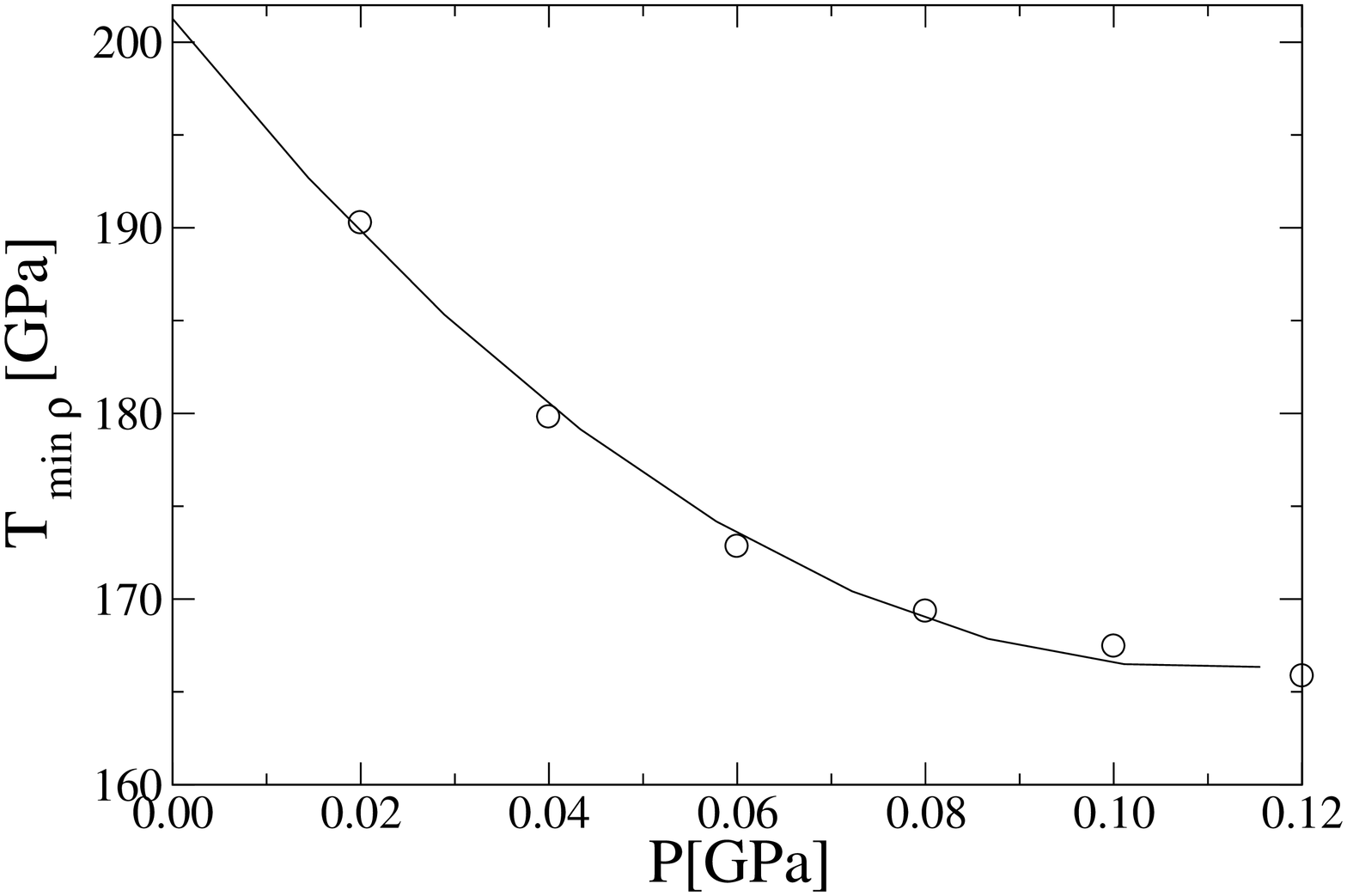}
\includegraphics[width=6cm,angle=-90]{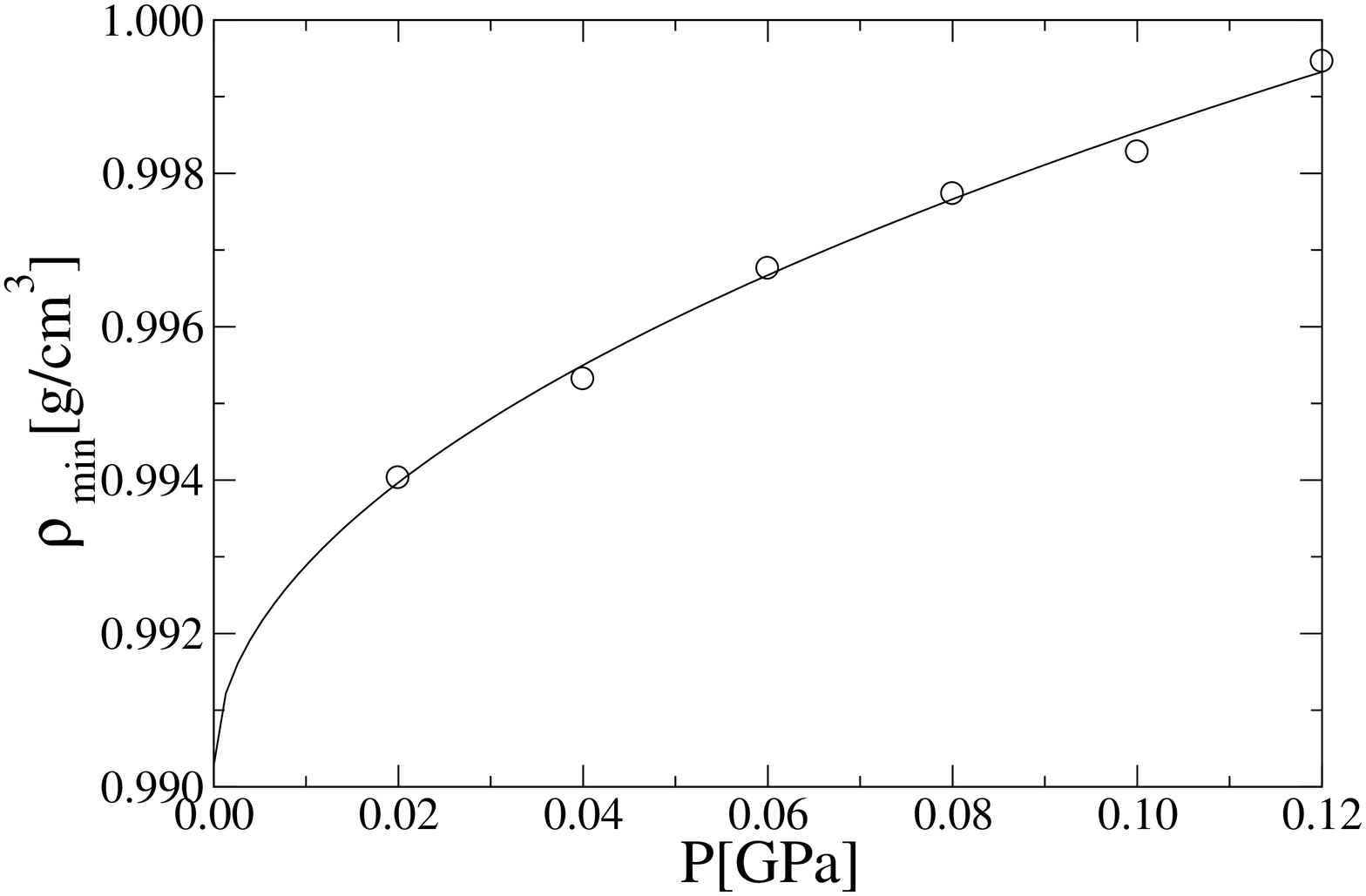}
\caption{Density minima for the coarse-grained model of a  water
  monolayer confined  between hydrophobic walls.
(Left panel) The locus of temperature $T_{\rm{min~\rho}}$ (circles)
of minimum density as a function of pressures $P$
follows approximately a quadratic curve (continuous line) in the $P$-$T$ plane,
that extrapolates to  about 201~K for atmospheric pressure.
(Right panel) The density minima $\rho_{\rm{min}}$  (circles) display
an approximate square-root
dependence on pressure $P$, whose extrapolation
for atmospheric pressure is about
$0.99$~g/cm$^3$.
Model parameters as in \ref{phase-diagram}.
}
\label{min-rho}
\end{figure}

\begin{figure}
\includegraphics[width=12cm,angle=-90]{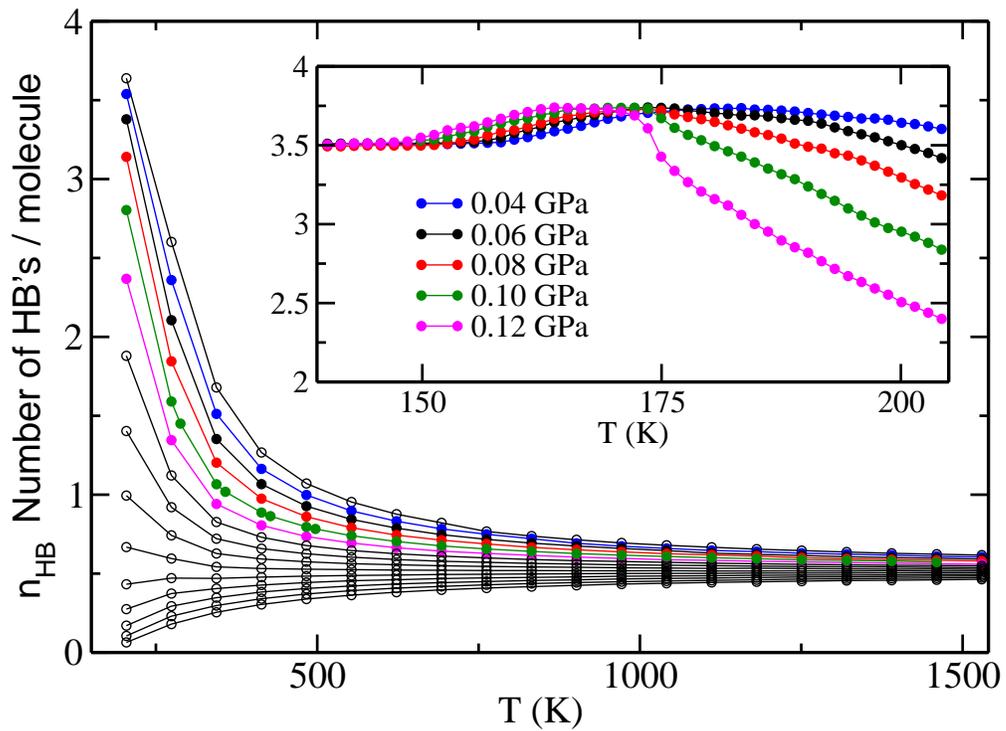}
\caption{Number of hydrogen bonds $n_{HB}$ to which a molecule
  participate, as a function of
  temperature $T$ for
  pressures $P$ from (top to bottom) $0.02$ to $0.3$~GPa
in increments of 0.02~GPa.
Inset: At low $T$ for $P$ from (top to bottom at 200~K) 0.04 to 0.12~GPa
we find that $n_{HB}$ reaches a maximum value of $\sim 3.75$
at temperatures that
coincide, within error bars, with those of the density minima
$\rho_{\rm{ min}}$ at the
same pressure. Model parameters as in \ref{phase-diagram}.}
\label{nhb}
\end{figure}

\begin{figure}
\includegraphics[width=12cm,angle=-90]{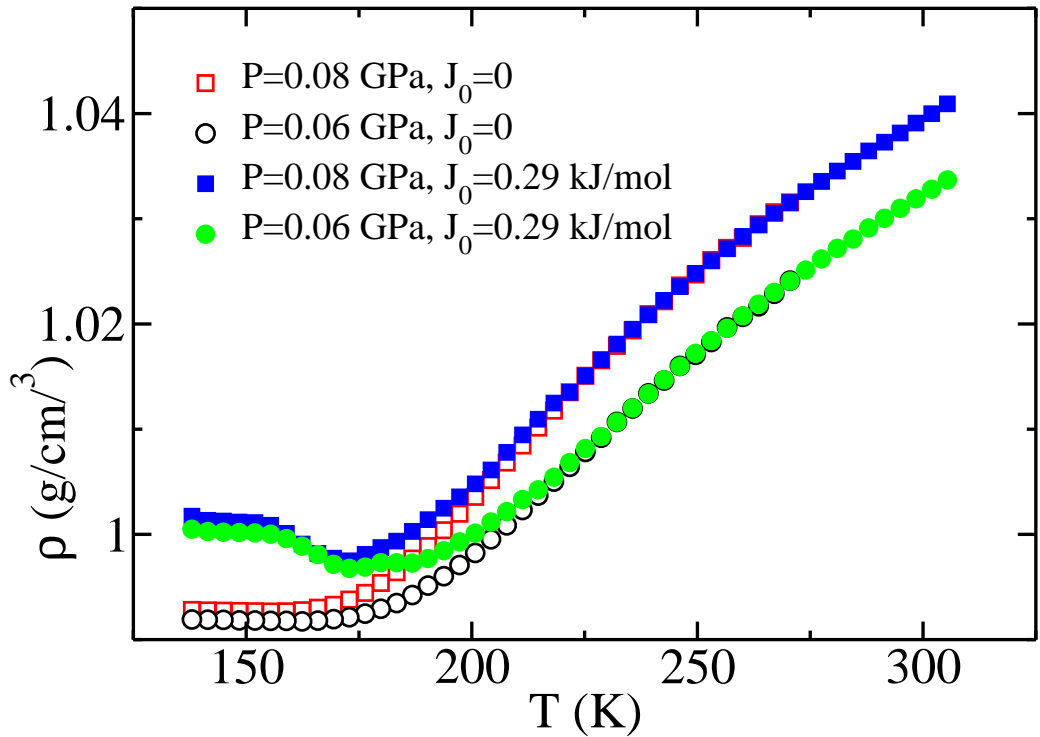}
\caption{Comparison of the density behavior at low $T$ for the model
  with parameters as in \ref{phase-diagram}
for $P=0.06$~GPa (full squares) and $P=0.08$~GPa
(full circles), and with
$J_\sigma=0$, and the other parameters unchanged,
for the same pressures (open squares and open circles,
respectively). We find that for $J_\sigma=0$ the density minima is not
detectable within our resolution.}
\label{compare-rho-min}
\end{figure}

\begin{figure}
\includegraphics[width=12cm,angle=0]{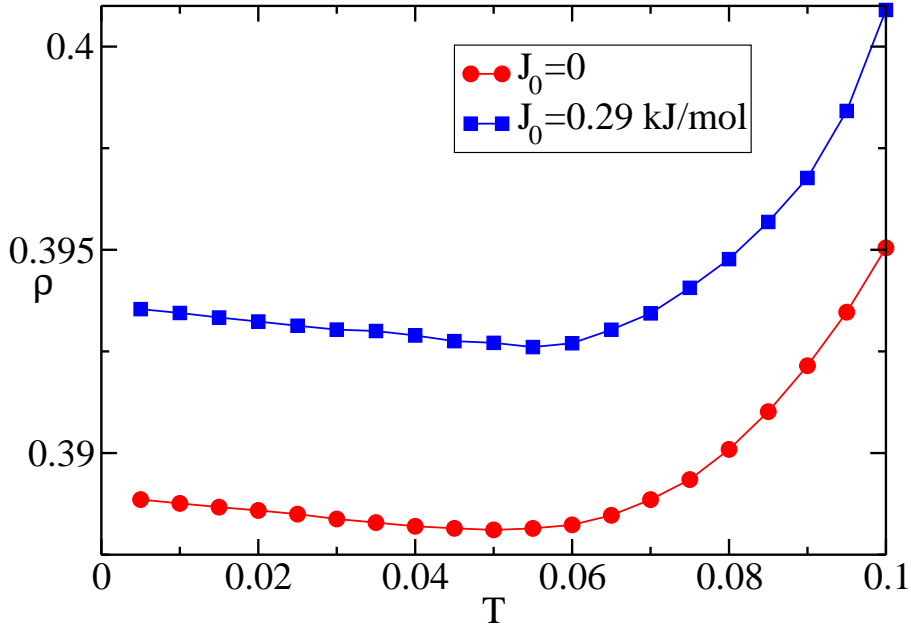}
\caption{As in \ref{compare-rho-min} but for $P=0.02$~GPa,
  with parameters as in \ref{phase-diagram} (squares) and with
  $J_\sigma=0$ (circles).
The density
  minimum occurs also when $J_\sigma=0$.}
\label{compare-low-P}
\end{figure}

Although for $T<T_o(P)$ our MC simulations become subdiffusive, we
find that we can equilibrate the HBs dynamics within our
simulation time
for temperatures as low as 190~K
at 0.02~GPa, or 163~K at 0.12~GPa. Specifically, we find that the
relaxation time of the bonding indices $\sigma_{ij}$, related to the
formation of the HBs, is of the order of $4$~ns for these
state points, while it exceeds our simulation time at lower $T$, e. g. at
about 170~K for $P=0.02$~GPa  \cite{FDLS2009}.

In the region of state points where we can equilibrate the
system, but close to the lowest well-equilibrated temperature,
we observe a minimum in density along isobars with $P\leq 0.12$~GPa
(\ref{density}).
This result resembles the experimental
density minimum
for water confined in a nanoporous silica matrix MCM-41 with a
pore diameter of 1.4~nm found by
Mallamace et al. \cite{Mallamace-density-min}.

From our simulations for $P\geq 0.02$~GPa, at atmospheric pressure  we
extract a density minimum of about
$0.99$~g/cm$^3$ at about 201~K (\ref{min-rho}) not too far from the
experimental value
$0.940\pm 0.003$~g/cm$^3$
at about $203\pm 5$~K and atmospheric external pressure
\cite{Mallamace-density-min}.
Although our simple quadratic extrapolation
predicts a value for $\rho_{\rm{min}}$ that is  larger than the
experimental,
our data give an extrapolated  $T_{\rm{min~\rho}}$
at atmospheric pressure consistent with the
results of the experiments \cite{Mallamace-density-min} and comparable
to those from MD simulations of TIP5P-E water
\cite{PhysRevLett.94.217802}.

It must be noted, however, that the experimental results for confined
water are controversial \cite{Mancinelli10}. Nevertheless, the
controversy is mainly about the experimental measurement of
the effect, and not about the effect,
because it has been observed that the
existence of a density minimum in water is a necessary consequence of
the existence of the low-$P$ branch of the TMD line
\cite{Peter-H-Poole:2005ad}.
In particular, it has been proposed that the locus of $\rho_{\rm min}$
corresponds to saturation, or the maximal ordering, of a network of
water molecules with a random tetrahedral local arrangement
\cite{Peter-H-Poole:2005ad}.

Our results, however, lead to a different explanation. We calculate the
number of HBs $n_{HB}$ in which a molecule participates,
defined as $n_{HB}\equiv 2N_{HB}/N$ from \ref{N_HB} in such a way as to
have four as maximum value for each molecule of the coarse-grained monolayer.
First, we observe that
our simulation results for the monolayer are consistent with
experimental data for bulk water, with  $n_{HB}\simeq 0.45$
at $P=0.25$~GPa and $T\simeq 670$~K, and with $n_{HB}\simeq 2.2$ at
the lowest $P$ at about ambient $T$, as reported in Ref.~\cite{bernabei:021505}
(\ref{nhb}).

Next, we find that for  $P\leq 0.12$~GPa
 the quantity reaches a maximum value of $n_{HB}\simeq 3.75$
at temperatures that
coincide, within error bars, with the temperatures $T_{\rm{min~\rho}}$
of the density minima $\rho_{\rm{ min}}$ at the
same pressure, and $n_{HB}$ decreases to $\sim 3.5$ below
$T_{\rm{min~\rho}}$. Therefore, $N_{HB}$ decreases and for \ref{Vw} the density
increases.

To understand why $n_{HB}$ decreases below its maximum $n_{HB}\simeq
3.75$  at low $T$, we compare two cases corresponding to two different
set of parameters of the model.
The first as in \ref{phase-diagram},
and the second, with  $J_\sigma=0$ and the other parameters
unchanged, corresponding to the LLCP and the SF scenario,
respectively \cite{Stokely2010}.
By comparing the low-$T$ behavior of density for the two cases at
intermediate $P$, we find that if $J_\sigma=0$, i.e. the hydrogen
bond is not cooperative, then the density minima is undetectable
within our resolution (\ref{compare-rho-min}).

However, at lower $P$ we find that both sets of parameters give a
detectable density minimum (\ref{compare-low-P}). Therefore, the
cooperative term of the HB interaction is not essential for
the occurrence of the density minima, but it emphasizes the minima at
intermediate pressures. From this observation we understand
that the explanation proposed by Poole at
al. \cite{Peter-H-Poole:2005ad} can be applied to the case with
$J_\sigma=0$, in which the HB interaction does not include a
cooperative (many body) term and the SF scenario is reproduced. In
this case, the density increases for decreasing $T$ at very low $T$,
when all the possible HBs have been formed, generating
regions of mismatching tetrahedral local order. A
decrease of $T$ induces a small reduction of free volume per
molecule, and a consequent small increase of density.

Instead, when $J_\sigma>0$, the cooperative interaction  in \ref{coop} induces the
breaking of HBs for decreasing $T$, to allow the reorientation of a molecule and
a  better matching of local tetrahedral order at low $T$. As a
consequence, the number $n_{HB}$
decreases from 3.75 to 3.5 inducing a large density increase.
The high energy cost of this local rearrangement, i.e. the energy
needed to break a HB, is at the origin of the high
energy barrier for the process and the
large increase of correlation time for the dynamics of
the HB network in the vicinity of the locus of density minima.

\subsection{Relation of diffusion anomaly with different scenarios}

Several authors relate the
diffusion anomaly in water to the presence of defects in the network
of HBs. Here we
show that the anomalous behavior of $D_{\parallel}$ in the
coarse-grained model is not related to the many-body component of the
HB interaction and, therefore,
to the possible occurrence of the LLCP.  The
anomaly is, instead, due  to the anticorrelation between volume and
entropy,  or to volume and energy, rooted to HB
formation.

\begin{figure}
\includegraphics[width=12cm,angle=-90]{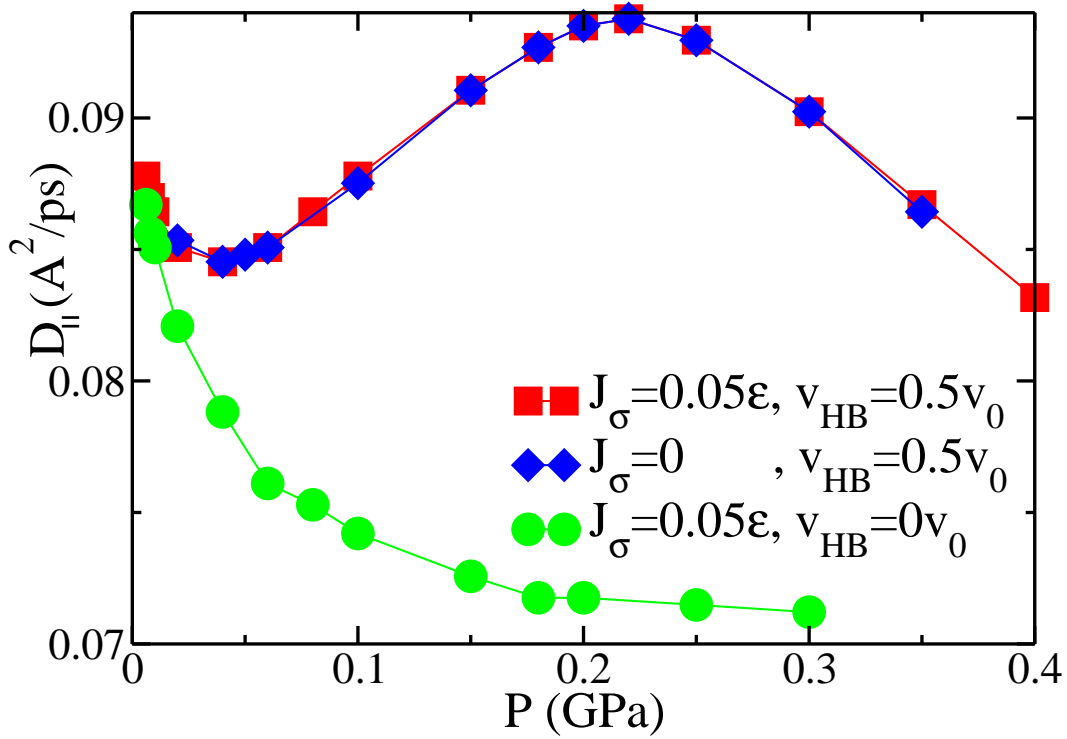}
\caption{Effect of parameters $v_{\rm HB}$ and $J_\sigma$ on the
  diffusion anomaly. We diffusion minima and maxima for both
  $J_\sigma>0$ and  $J_\sigma=0$. Instead, the non-monotonic behavior of
  $D_{\parallel}$ disappears for $v_{\rm HB}=0$.}
\label{vh0}
\end{figure}

We consider three different realizations of the coarse-grained
model (\ref{vh0}). The first corresponds to the case presented in the
previous section, with the parameters as in \ref{phase-diagram}.
For the second, we set
$J_\sigma=0$, leaving the other parameters
unchanged. This
case reproduces the SF scenario\cite{Sastry1996},  where density
maxima occur and which has been
shown to correspond to the vanishing-$T$ limit of the LLCP
\cite{Stokely2010}.
Our simulations show that the
occurrence of the anomaly of $D_{\parallel}$ is unaffected by this
change of parameters (\ref{vh0}).
Therefore, the absence of cooperativity
in the HB dynamics is not relevant for the
occurrence of both density maxima and diffusion anomaly.

This can be understood for the clear separation between the
temperature range at which the anomaly of $D_{\parallel}$ occurs and
the temperature range at which water becomes subdiffusive. Only the latter
regime corresponds to the temperature range for which the
cooperativity has a strong influence on the dynamics, while it has no
major dynamic effect at higher $T$.

Next, we set $v_{HB}=0$ and leave the other parameters as in \ref{phase-diagram}.
This case would correspond to cooperatively bonding liquid
with no density anomaly, i.e. with no anticorrelation between volume
and entropy or volume and energy. In this case the change in the
behavior of $D_{\parallel}$ is striking (\ref{vh0}). The system has no
diffusion anomaly, with $D_{\parallel}$ that decreases monotonically
for increasing $P$ as in normal liquids.

Therefore, this result clarifies that the anomalous volume behavior
due to the HB formation is directly related to the
anomalous diffusion behavior. This conclusion, and
the previous observation
that the many-body component of the HB interaction
is not relevant in determining
these anomalies, leads us to investigate how the volume available
for diffusion, and the number of HBs are related to
$D_{\parallel}$, as discussed elsewhere \cite{DLSF2010}.

\section{Conclusions}

We study by Monte Carlo simulations a coarse-grained model for a water monolayer confined
between hydrophobic walls.
We consider a separation between walls about
$h=0.5$~nm that  inhibits the formation of ice
\cite{PhysRevLett.91.025502}.

The model includes isotropic, directional (covalent) and many-body
(cooperative) components of the HBs.
Thanks to its coarse-graining, the model
allows to study water in extreme conditions and, also, to check how
each of the HB component affects its properties.
Moreover, it makes possible to perform mean field calculations to
compare with simulations results.

We find gas and liquid phases, separated by a boundary line of first--order phase
transitions ending in a critical point occurring at higher pressure and
temperature, consistent with other models for hydrophobically confined water
\cite{Gallo07}. We study the diffusion constant parallel to the
walls $D_{\parallel}$ and find that it displays a line of maxima and a
line of minima at constant $T$, as seen in similar confinement for
other models \cite{Kumar2005}. Our analysis allows us to conclude that the
anomalous $D_{\parallel}$ is a consequence of the
anomalous volume behavior due to HB formation. 
In particular, the positive correlation between entropy and density,
or energy and density, due
to hydrogen bonding is the key element for the diffusion anomaly.
It is worth reminding here that a similar result has been found also
for potentials with isotropic interactions and water-like anomalies
\cite{vilaseca:084507,Errington06,oliveira08}
when the isothermal density dependence of the excess
entropy, which is related to the total isothermal entropy by a linear
function of the logarithm of the density, is considered.

The difference with the present analysis is, nevertheless, threefold. 
First, for these isotropic potentials the HDL phase has less entropy
than the LDL phase, implying a positive slope in the $P$-$T$ plane
for the liquid-liquid phase coexistence line as a consequence of the
Clausius-Clapeyron equation. Instead, for water and the present model the HDL phase has
more entropy than the LDL phase, hence the liquid-liquid phase
transition has a negative slope in the $P$-$T$ plane.

Second, here we show that by setting the parameter that controls the
increase of volume for HB formation, hence the positive correlation of
density with entropy and energy, the anomalous diffusion behavior
vanishes. Instead,
for the isotropic potentials the vanishing of the anomalous diffusion
behavior is controlled by the softness of the soft-core repulsion of
the potential \cite{vilaseca:084507,Vilaseca2011419}. A direct relation between
these two results is not straightforward and could be interesting to
investigate.

Third, the present result does not exclude that the key element for
diffusion anomaly is the positive correlation of density and energy,
instead of entropy. While in water and the present model the LDL phase
has lower energy and entropy than the HDL, in the isotropic potentials
with water-like anomalies the LDL phase has lower energy but higher
entropy than the HDL. These considerations support the idea that the
mechanism of anomalies in isotropic potentials is different from that
of water \cite{Franzese:2001ea}.

Interestingly, here we also observe  that the lines of constant $D_{\parallel}$
 resemble the melting line of bulk water.
At low temperatures, we find the locus of density maxima, which marks
another well-known water anomaly. We discuss how this locus is related
to the locus of expansivity minima.

At lower $T$, we find subdiffusive behavior $\Delta
r^2\sim t^\alpha$, as seen in experiments of hydration water
\cite{Settles-Doster1996} and simulations of confined water
\cite{Gallo:2003vn,Gallo2010,Bizzarri:1996uq,Bizzarri:1996fk,Rocchi:1998kx,Oleinikova:2007bh}.
Our results are well described by $\alpha=0.7$, between 0.96 \cite{Bizzarri:1996fk}
and $0.45\pm 0.05$ \cite{Gallo:2003vn} of previous calculations.
Previous works proposed that subdiffusion is a consequence of
the heterogeneities in water-interface interaction. Here this
rationale does not apply and we relate the subdiffusion to the
increase of correlation among water molecules at low $T$ due to the
full development of the HB network.

By further decreasing $T$, we find density minima, as seen in
experiments \cite{Mallamace-density-min} and MD
simulations \cite{PhysRevLett.94.217802,Peter-H-Poole:2005ad}.
These minima occur within the subdiffusive part of the phase diagram,
therefore where translational motion is strongly hampered and glassy
behavior is incipient. Nevertheless, the HB network within
this region is still dynamically evolving, with increasing correlation
time \cite{FDLS2009}. In particular, the HB
correlation time is about 4~ns for the majority of the subdiffusive
region and increases, exceeding our simulation times of the order of
$15$~ms, only at about $P\leq 0.02$~GPa and $T\leq 170$~K.

Previous works related the density minima to the saturation of a network of
molecules with a random tetrahedral local arrangement
\cite{Peter-H-Poole:2005ad}.  However, this rationale apply to our
model only for the case in which the cooperative component of the hydrogen
bonds is zero. When the cooperative component is larger than zero, as
expected in real water \cite{Cooperativity}, our calculations show that the
minima are due to a reduction of the number of HBs,
as a consequence of the reorientation of molecules for a better
matching of local order. The high energy cost of this
rearrangement is the cause of the large slowing down of the HB
dynamics near the state points where the density minima occur.

\acknowledgement

FDLS acknowledges support from projects nos. FIS2009-08451 (MICINN)
and P07-FQM02725 (Junta de Andaluc\'ia). GF acknowledges support from
MICINN project n. FIS2009-10210 (co-financed FEDER).


\providecommand*\mcitethebibliography{\thebibliography}
\csname @ifundefined\endcsname{endmcitethebibliography}
  {\let\endmcitethebibliography\endthebibliography}{}

\end{document}